\def\j{{\mathbf j}}

\def\be{\begin{equation} }
\def\ee{\end{equation} }

\def\n{\noindent }
\documentclass[12pt,a4paper,epsfig]{iopart}
\usepackage{epsfig}
\begin{document}
\title[Optical conductivity in disordered alloys]
{Optical conductivity in disordered alloys : an approach via the augmented space recursion} 
\author{Kartick Tarafder\footnote{E-mail : kartick@bose.res.in} and Abhijit Mookerjee\footnote{E-mail : abhijit@bose.res.in}}
\address{S.N. Bose National Centre for Basic Sciences,
JD Block, Sector III, Salt Lake City, Kolkata 700098, India}
\begin{abstract}
We present here a calculation of the configuration averaged optical conductivity
of random binary alloys CuAu and AgAu. Our formulation is based on the augmented
 space formalism proposed by  Mookerjee [J. Phys. C : Solid State Phys. {\bf 6} 
1340 (1973)] and the optical conductivity is obtained directly through a recursive procedure suggested by Viswanath and M\"uller [``The user friendly recursion method", Troisieme Cycle de la Physique, en Suisse Romande (1993)].
\end{abstract}
\pacs{71.23.-k}
%%%%%%%%%%%%%%%%%%%%%%%%%%%%%%%%%%%%%%%%%%%%%%%%%%%
\section{Introduction}
In an earlier work \cite{km1} we had developed a methodology for the calculation of configuration averaged optical conductivity of a disordered alloy based on the augmented space formalsim developed by one of us \cite{am1}. The method was based on the Kubo-Greenwood formula and we had shown that disorder scattering renormalizes
both the Green functions and the current terms.

In linear response theory, at zero temperature, the generalized susceptibility of a disordered alloy is given by the Kubo formula : 

\[ \langle \j^\mu(t)\rangle = \int_{-\infty}^\infty \chi^{\mu\nu}(t-t')\ A^\nu(t') \]

\n where, $A^\nu(t)$ is the vector potential, and 

\[ \chi^{\mu\nu}(t-t') = (i/\hbar)\ \Theta(t-t')\ \langle[\j^\mu(t) \j^\nu(t')]\rangle \]

\n {\bf j}$^\mu$ is the current operator and $\Theta$ is the heaviside step function.
If the underlying lattice has cubic symmetry, $\chi^{\mu\nu}$ = $\chi\ \delta_{\mu\nu}$.
The fluctuation dissipation theorem then relates the imaginary part of the Laplace
transform of the generalized  susceptibility to the Laplace transform of a correlation
function :

\[
\chi^{\prime\prime}(\omega) = (1/2\hbar) \left(1-\exp\{-\beta\hbar\omega\}\right)S(\omega) \]
\n where,

\begin{equation}
 S(\omega) = \int_0^\infty\ dt \ \exp\{i(\omega + i\delta)t\}\ \Tr \left(\rule{0mm}{4mm} \j^\mu(t)\ \j^\mu(0)\ \right)
\end{equation}

Since the response function is independent of the direction label $\mu$ for cubic symmetry, in the
following we shall drop this symbol. In case of other symmetries we have to generalize our results
for different directions.
Our goal will be, given a quantum ``Hamiltonian" {\bf H} to obtain the
correlation function,

\[
S(t) = \langle \phi\vert \mathbf{j}(t)\mathbf{j}(0)\vert\phi\rangle
\]

We shall now determine the correlation directly via the recursion method as described by Viswnath and M\"uller\cite{gb}. In order to simplify the expressions for the dynamical quantities as produced by the Hamiltonian, we consider henceforth the modified Hamiltonian $\bar{\bf H}\ =\ {\bf H} - E_0{\bf I}$, whose band energy is shifted to zero. Let

\[ \langle \psi(t)\vert\ =\ \langle\phi\vert\ \mathbf{j}(t) \]

\n The time evolution of this {\sl bra} is goverened by the Schr\"odinger equation

\begin{equation}
 -i\ \frac{d}{dt}\left\{\rule{0mm}{4mm} \langle\psi(t)\vert\right\}\ =\ \langle\psi(t)\vert\bar{\mathbf {H}} 
\label{eq2}
\end{equation}

We shall now generate an orthogonal basis $\{\langle f_k\vert\}$ for representation of equation~(\ref{eq2}). 
We shall do this in the following way :

\begin{enumerate}
\item[(i)]We begin with initial conditions : \[ \langle f_{-1}\vert = 0\quad ;\quad \langle f_0\vert = \langle \phi\vert{\bf j}(0)\]
\item[(ii)] We now generate the new basis members by a three term recurrance
relationship :
\[ \langle f_{k+1}\vert\ = \ \langle f_k\vert \bar{\mathbf H}\ - \ \langle f_k\vert \alpha_k \ -\ \langle f_{k-1}\vert \beta_k^2\quad\quad \mathrm {k=0,1,2}\ldots\]

\n where, \[ \alpha_k = \frac{\langle f_k \vert\bar{\bf H}\vert f_k\rangle}{\langle f_k \vert f_k\rangle}\quad\quad \beta_k^2= \frac{\langle f_k\vert f_k\rangle}{\langle f_{k-1}\vert f_{k-1}\rangle } \]
\end{enumerate}

\n We now expand the bra $\langle \psi(t)\vert $ in this orthogonal basis :

\begin{equation}
 \langle \psi(t)\vert \ =\ \sum_{k=0}^\infty\ \langle f_k\vert\ D_k(t) 
\label{eq3}\end{equation}

\n Substituting equation (\ref{eq3}) into equation (\ref{eq2})  and 
using orthogonality of the basis, we get~:

\begin{equation}
-i \dot{D}_k(t) \ =\ D_{k-1}(t)+ \alpha_k\ D_k(t) + \beta^2_{k+1}\ D_{k+1}(t)
\label{eq4}\end{equation}

\noindent with $D_{-1}(t) = 0$ and $D_k (0)=\delta_{k,0}$. We shall now show that the
pair of sequences generated by us, namely, $\{\alpha_k\}$ and $\{\beta_k^2\}$
are enough for us to generate the correlation function.
We note first that :

\begin{equation}
 D_0(t) = \langle\psi(t)\vert f_0\rangle
 = S(t)
\end{equation}

\n Let us define the Laplace transform :

\[ d_k(z) \ =\ \int_0^\infty\ dt\ \exp{(-izt)}\ D_k(t) \]

\n Putting this back in equation(\ref{eq4}) we get :

\begin{equation}
(z-\alpha_k)\ d_k(z)\ -\  i\delta_{k,0}\ =\ d_{k-1}(z)+\beta^2_{k+1}\ d_{k+1}(z)\quad\quad\mbox{k=0,1,2}\ldots
\label{eq5}\end{equation}

\n This set of equations can be solved for $d_0(z)$ as a continued fraction 
representation~:

\begin{equation}
d_0(z) \ =\ \frac{i}{\displaystyle z-\alpha_0-\frac{\beta^2_1}{
\displaystyle{z-\alpha_1- \frac{\beta^2_2}{\displaystyle z-\alpha_2 - \ldots}}}}
\end{equation}

The structure function, which is the Laplace transform of the correlation function can then be obtained from the above :

\begin{equation}
 S(\omega)\ =\ \lim_{\delta\rightarrow 0}\ 2\ \Re e\ d_0(\omega+i\delta) 
\label{eq6}\end{equation}

\section{Augmented space formulation}

Augmented space method was first proposed by one of us \cite{am1} as a feasible
technique for carrying out configuration averaging in disordered systems. In
particular, it has been applied extensively for electronic structure of disordered alloys. The readers are referred to the article \cite{tf1} for details. Here we shall discuss the salient features of this method for application to configuration averages of correlation functions.

The problem addressed involves the configuration averaging of a function of many
independent random variables, e.g. $\ll f(\{n_R\})\gg$ where $R$ labels a set
of lattice points. The first step is to associate with each random variable
$n_R$ an operator ${\bf M}^R$ such that the spectral density of this operator is the probability density of the random variable :

\[ p(n_R) = -(1/\pi) \langle \uparrow\vert (n_R\mathbf{I} - {\mathbf M}^R)^{-1}\vert \uparrow\rangle \]

For example, if $n_R$ takes the values 0 and 1 with probabilities $x$ and $y=1-x$
then the operator {\bf M}$^R$ is of rank 2 and acts on the configuration space of $n_R$, $\phi^R$, spanned by the eigenstates $\vert 0\rangle$ and $\vert 1\rangle$ of {\bf M}$^R$. The
state $\vert\uparrow\rangle = \sqrt{x}\vert 0\rangle + \sqrt{y}\vert 1\rangle$
and the representation of {\bf M}$^R$ in the basis $\{\vert \uparrow\rangle,
\vert\downarrow\rangle = \sqrt{y}\vert 0\rangle -\sqrt{x}\vert 1\rangle \}$
is

\[ \left( \begin{array}{cc}
           x & \sqrt{xy} \\
  \sqrt{xy} & y \end{array} \right)\]

The augmented space theorem then states that :

\begin{equation}
\ll f(\{n_R\}) \gg = \langle \{\emptyset\} \vert \tilde{\mathbf f}(\{{\mathbf M}^R\}) \vert \{\emptyset\} \rangle 
\end{equation}

\n where $\tilde{\mathbf f}$ is an operator of rank $2^N$ obtained by replacing each $n_R$ by {\bf M}$^R$. This operator acts on the configuration space of the set of random variables, $\Phi$ = $\prod^\otimes \phi^R$. A state in this space is given by a pattern of configurations at each site, i.e. sequences like $\{\uparrow\uparrow\downarrow\ldots \}= \{{\cal C}\}$ which can be uniquely represented by the sequence of sites where we have (say) a $\downarrow$ configuration. This sequence is called a
{\sl cardinality sequence}. The empty cardinality sequence, $\{\emptyset\}$, is the pattern of $\uparrow$-s everywhere.

If we go back to equations (2)-(5), we notice that for a disordered binary
alloy,  $S(t)= S[\bar{\mathbf H}(\{n_R\})]$. The augmented space theorem then
states that :

\[\fl \ll S(t)\gg = \ll \langle \phi\vert {\mathbf j}(t){\mathbf j}(0)\vert\phi\rangle \gg = \langle \phi\otimes \{\emptyset\}\vert \tilde{\mathbf j}(t)\tilde{\mathbf j}(0)\vert\phi\otimes\{\emptyset\}\rangle = S[\widetilde{\mathbf H}(\{{\mathbf M}^R\})] \]

\n where the augmented space Hamiltonian and the current operators are constructed by replacing every random variable $n_R$ by the corresponding operator {\bf M}$^R$.

\n The recursion may now be modified step by step in the full augmented space :

\[ \langle \psi(t)\vert\ =\ \langle\phi\otimes\{\emptyset\}\vert\ \tilde{\mathbf{j}}(t) \]

\n The time evolution of this {\sl bra} is goverened by the Schr\"odinger equation

\begin{equation}
 -i\ \frac{d}{dt}\left\{\rule{0mm}{4mm} \langle\psi(t)\vert\right\}\ =\ \langle\psi(t)\vert\widetilde{\mathbf {H}} 
\label{eq2a}
\end{equation}

As before, we shall generate the orthogonal basis $\{\langle f_k\vert\}$ for representation of equation (\ref{eq2a}) :

\begin{enumerate}
\item[(i)]We begin with initial conditions : \[ \langle f_{-1}\vert = 0\quad ;\quad \langle f_0\vert = \langle \phi\otimes\{\emptyset\}\vert\tilde{\bf j}(0)\]
\item[(ii)] The new basis members are generated by a three term recurrance
relationship :
\[ \langle f_{k+1}\vert\ = \ \langle f_k\vert \widetilde{\mathbf H}\ - \ \langle f_k\vert\ \tilde{\alpha}_k \ -\ \langle f_{k-1}\vert\ \tilde{\beta}_k^2\quad\quad \mathrm {k=0,1,2}\ldots\]

\n where, \begin{equation}
 \tilde{\alpha}_k = \frac{\langle f_k \vert\widetilde{\bf H}\vert f_k\rangle}{\langle f_k \vert f_k\rangle}\quad\quad \tilde{\beta}_k^2= \frac{\langle f_k\vert f_k\rangle}{\langle f_{k-1}\vert f_{k-1}\rangle } 
\end{equation}
\end{enumerate}

\n We now expand the bra $\langle \psi(t)\vert $ in this orthogonal basis :

\[
 \langle \psi(t)\otimes\{\emptyset\}\vert \ =\ \sum_{k=0}^\infty\ \langle f_k\vert\ \widetilde{D}_k(t) 
\]

\n Continuing exactly as in the last section we get,
\begin{equation}
 \widetilde{D}_0(t) = \langle\psi(t)\otimes \{\emptyset\}\vert f_0\rangle
 = \ll S(t)\gg
\end{equation}

\n Taking Laplace transforms and using the three term recurrance,

\begin{equation}
\tilde{d}_0(z) \ =\ \frac{i}{\displaystyle z-\tilde{\alpha}_0-\frac{\tilde{\beta}^2_1}{
\displaystyle{z-\tilde{\alpha}_1- \frac{\tilde{\beta}^2_2}{\displaystyle z-\tilde{\alpha}_2 - \ldots}}}}
\end{equation}

The configuration averaged structure function, which is the Laplace transform of the averaged correlation function can then be obtained from the above :

\begin{equation}
\ll S(\omega)\gg \ =\ \lim_{\delta\rightarrow 0}\ 2\ \Re e\ \tilde{d}_0(\omega+i\delta) 
\label{eq6a}\end{equation}

The equations (11)-(14) will form the basis of our calculation of the configuration averaged correlation function. 

\section{The Hamiltonian and current operators in augmented space}

Let us begin with a Hamiltonian for a random binary alloy represented in the basis of a tight-binding linear muffin-tin orbitals method (TB-LMTO) $\{\vert\chi_R\rangle\}$ \cite{aj}. For the sake of notational clarity we have suppressed the angular momentum index $L$. In what follows, we may consider $R$ to be a composite index $RL$.

\[ 
{\mathbf H}\ =\ \sum_R C_R\ {\mathbf P}_R + \sum_R\sum_{R'} \ \Delta_R^{1/2}\ S_{RR'}\ \Delta_{R'}^{1/2}\ {\mathbf T}_{RR'}
\]

\n {\bf P}$_R$ and {\bf T}$_{RR'}$ are projection and transfer operators on the space spanned by the TB-LMTO basis. $C_R$ and $\Delta_R^{1/2}$ are the TB-LMTO potential parameters and $S_{RR'}$ the structure matrix. For a binary alloy the former are random and may be given by

\begin{eqnarray*}
C_R = C_A\ n_R + C_B\ (1-n_R) = C_B + (C_A-C_B)\ n_R \\
\Delta_R^{1/2} = \Delta^{1/2}_A\ n_R + \Delta^{1/2}_B\ (1-n_R) = \Delta^{1/2}_B + (\Delta^{1/2}_A-\Delta^{1/2}_B)\ n_R 
\end{eqnarray*}

\n Here the random variables $n_R$ takes values 0 and 1 with probabilities $x$ and $y$, as before.
The augmented space theorem then builds up the augmented Hamiltonian by
replacing the binary random variables $\{n_R\}$ by the corresponding
operators $\{{\mathbf M}^R\}$ described in the earlier section.

\begin{eqnarray}
\fl \widetilde{\mathbf C}_R = \ll C \gg {\mathbf I} + (y-x)(C_A-C_B)\
{\mathbf P}^\downarrow_R + \sqrt{xy}(C_A-C_B)\ {\mathbf T}^{\downarrow\uparrow}_R \nonumber\\
\fl \widetilde{\Delta}_R^{1/2} = \ll \Delta^{1/2} \gg {\mathbf I} + (y-x)(\Delta^{1/2}_A-\Delta^{1/2}_B)\ {\mathbf P}^\downarrow_R + \sqrt{xy}(\Delta^{1/2}_A-\Delta^{1/2}_B)\ {\mathbf T}^{\downarrow\uparrow}_R 
\end{eqnarray}

\n where the configuration space operators are :

\begin{eqnarray*}
{\mathbf P}^\downarrow_R \vert \{{\cal C}\}\rangle =
\delta\left(R\in\{{\cal C}\}\right)\ \vert \{{\cal C}\}\rangle \\
{\mathbf T}^{\downarrow\uparrow}_R \vert \{{\cal C}\}\rangle =
\vert \{{\cal C}\}\pm R \rangle
\end{eqnarray*}

\n The augmented space Hamiltonian is then given by :

\begin{equation}
\widetilde{\mathbf H} = \sum_R\ \widetilde{\mathbf C}_R\ \otimes\  {\mathbf P}_R\
+\ \sum_R\sum_{R'}\ \widetilde{\Delta}^{1/2}_R \ S_{RR'} \widetilde{\Delta}^{1/2}_{R'}\ \otimes\ {\mathbf T}_{RR'}
\end{equation}

\n Next we look at the expression for the current operator in the TB-LMTO basis :

\[
\vert\chi_R\rangle=\vert\phi_R\rangle +\sum_{R^{\prime}}h_{RR\prime}\vert\dot{\phi}_{R^{\prime}}\rangle
\]

\n The dot refers to derivative with respect to energy.
In this basis, the matrix elements of the current operator can be written as

\begin{eqnarray}
\fl J^\mu_{RR^{\prime}}=\langle\chi_{R^{\prime}}\vert {\mathbf j}^\mu\vert\chi_{R}\rangle \nonumber\\
\fl \phantom{J^\mu_{RR^{\prime}}}= e\left[V^{(1),\mu}_{RR^{\prime}}\ \delta_{RR^{\prime}}+\sum_{R^{\prime\prime}}V^{(2),\mu}_{RR^{\prime\prime}}\ h_{R^{\prime\prime}R^{\prime}}+
 \sum_{R^{\prime\prime}}h_{RR^{\prime\prime}}\ V^{(3),\mu}_{R^{\prime\prime}R{\prime}}+\sum_{R^{\prime\prime}}\sum_{R^{\prime\prime\prime}}h_{RR{\prime\prime\prime}}\ V^{(4),\mu}_{R^{\prime\prime\prime}R{\prime\prime}}\ h_{R^{\prime\prime}R^{\prime}}\right]\nonumber\\
\end{eqnarray}

\n where

\begin{eqnarray*}
 V^{(1),\mu}_{RR^{\prime}}=\langle\phi_{R^{\prime}}\vert {\mathbf v}^\mu\vert\phi_{R}\rangle \ ;\ 
 V^{(2),\mu}_{RR^{\prime}}=\langle\dot{\phi}_{R^{\prime}}\vert {\mathbf v}^\mu\vert\phi_{R}\rangle \\ 
 V^{(3),\mu}_{RR^{\prime}}=\langle\phi_{R^{\prime}}\vert {\mathbf v}^\mu\vert\dot{\phi}_{R}\rangle \ ;\ 
 V^{(4),\mu}_{RR^{\prime}}=\langle\dot{\phi}_{R^{\prime}}\vert{\mathbf v}^\mu\vert\dot{\phi}_{R}\rangle
\end{eqnarray*}

The technique for calculating these matrix elements has been described in detailed by Hobbs \etal \cite{hobbs}. We have also used this technique in our earlier
paper \cite{km1} and we shall use it here as well. The readers are referred to these two papers for details.

Ideally the next step would be  to calculate $J^\mu_{AA},J^\mu_{BB},J^\mu_{AB},J^\mu_{BA}$ as the current terms between two sites when they are occupied by atom pairs
AA, BB, AB and BA embedded in the disordered medium. A simpler first step would
be to obtain these current terms from the pure A and B and from the ordered AB alloy. In general, the current operator can be written as :

\be {\mathbf j}^\mu = \sum_R J^\mu(0)\ {\bf P}_R + \sum_R\sum_{R'}\ J^\mu(\chi)\ {\bf T}_{RR'}\ee

\n where $\chi = R-R'$.

In a disordered alloy, the current representations are random :

\begin{eqnarray*}
J^\mu(0)&=&J^\mu_{AA}(0)\ n_R+J^\mu_{BB}(0)\ (1-n_R) \qquad  \mathrm {and}\\
J^\mu(\chi)&=&J^\mu _{AA}(\chi)\ n_R\ n_{R'} +J^\mu _{BB}(\chi)\ (1-n_R)\ (1-n_{R'})\\
 &&+J^\mu _{AB}(\chi)\ n_R\ (1-n_{R'})+J\mu_{BA}(\chi)\ (1-n_R)\ n_{R'}
\end{eqnarray*}

Using the augmented space theorem,  we replace the random variable $n_R$ by an operator $M^R$ in the expression of $J^\mu(0),J^\mu(\chi)$ we get 
the current operator in the augmented space~:

\begin {eqnarray}
\fl\tilde{\mathbf j}^\mu=\langle J^\mu(0)\rangle\ {\mathbf I}\otimes {\mathbf P}_R +J^\mu_1(0)\ {\mathbf P}_R^{\downarrow}\otimes {\mathbf P}_R +J^\mu_2(0)\ {\mathbf T}_R^{\uparrow\downarrow}\otimes {\mathbf P}_R \nonumber\\
\fl\phantom{i} +\langle J^\mu(\chi)\rangle\  {\mathbf I}\otimes {\mathbf T}_{R,R'} +(y-x)\ J^\mu_1(\chi)\ {\mathbf P}_{R'}^{\downarrow}\otimes {\mathbf T}_{R,R'}
+(y-x)\ J^\mu_2(\chi)\ {\mathbf P}_R^{\downarrow}\otimes {\mathbf T}_{R,R'}\nonumber\\
\fl\phantom{i} +(y-x)^2\ J^\mu_3(\chi)\ {\mathbf P}_R^{\downarrow}\otimes {\mathbf P}_{R'}^{\downarrow}\otimes {\mathbf T}_{R,R'}
 +\sqrt{xy}\ \left\{\rule{0mm}{4mm}\right. J^\mu_1(\chi)\  {\mathbf T}_{R'}^{\uparrow\downarrow}\otimes {\mathbf T}_{R,R'}
+J^\mu_2(\chi){\mathbf T}_{R}^{\uparrow\downarrow}\otimes {\mathbf T}_{R,R'}\left.\rule{0mm}{4mm}\right\}\nonumber\\
\fl\phantom{i}+\sqrt{xy}(y-x)J^\mu_3(\chi)\ \left(\left[{\mathbf P}_R^{\downarrow}\otimes {\mathbf T}_{R'}^{\uparrow\downarrow}+{\mathbf T}_{R}^{\uparrow\downarrow}\otimes {\mathbf P}_{R'}^{\downarrow}\right]\otimes {\mathbf T}_{R,R'}\right)
+xy\ J^\mu_3(\chi)\ {\mathbf T}_{R}^{\uparrow\downarrow}\otimes{\mathbf T}_{R'}^{\uparrow\downarrow}\otimes {\mathbf T}_{R,R'}\nonumber\\
\end{eqnarray}
where 
\begin{eqnarray*}
\langle J^\mu(0)\rangle = xJ_{AA}^\mu(0)+yJ_{BB}^\mu(0)\qquad
J^\mu_1(0)=(y-x)[J_{AA}^\mu(0)-J_{BB}^\mu(0)]\\
J^\mu_2(0)=\sqrt{xy}[J_{AA}^\mu(0)-J_{BB}^\mu(0)]\\
\langle J(\chi)\rangle =x^2J_{AA}^\mu(\chi)+xy\ \left(J_{AB}^\mu(\chi)+J_{BA}^\mu(\chi)\right)+y^2J_{BB}^\mu(\chi)\\
J_1^\mu(\chi)=x\left(J_{AA}^\mu(\chi)-J_{AB}^\mu(\chi)\right)+y\left(J_{BA}^\mu(\chi)-J_{BB}^\mu(\chi)\right)\\
J_2^\mu(\chi)=x\left(J_{AA}^\mu(\chi)-J_{BA}^\mu(\chi)\right) +y\left(J_{AB}^\mu(\chi)-J_{BB}^\mu(\chi)\right)\\
J_3^\mu(\chi)=J_{AA}^\mu-J_{AB}^\mu-J_{BA}^\mu+J_{BB}^\mu
\end{eqnarray*}

This augmented current operator is used to construct the starting state of the recursion as described in the last section.

\section{Results and Discussion}
For application of our methodology we have chosen two alloy systems disordered : CuAu (50-50) and disordered AgAu (50-50). Our choice was governed by two considerations : first, we had studied these
alloys in an earlier work with a different approach \cite{km1} and it would be interesting to compare the two approaches ; and secondly, the fact that we have experimental data available for both these alloys and it would be interesting to compare our theoretical estimates with experiment.
We have begun our study with the self-consistent TB-LMTO-ASR calculation on AgAu and CuAu 50-50 alloys. We have
minimized the energy with respect to the variation in the average lattice constant for both the alloys.
We have calculated the current terms from the basis functions and potential parameters obtained from
a TB-LMTO calculation for the ordered 50-50 alloys. Ideally we should have embedded  AA, BB and AB pairs
in the disordered medium and obtained the current expressions from such a calculation. However, in our
earlier work \cite{km1} we have shown that effect of randomness in the current terms is small. This gives us confidence to proceed with these estimates.

To check the density of states for the disordered alloys we have used our TB-LMTO-ASR programme developed by us \cite{adc}. The density of states has been shown in our earlier work \cite{km1}. The main features of interest is that the filled $d$-states lie about 2 eV below the Fermi level for both these alloys. 

\begin{figure}
\centering
\epsfxsize=2in\epsfysize=2.5in\rotatebox{270}{\epsfbox{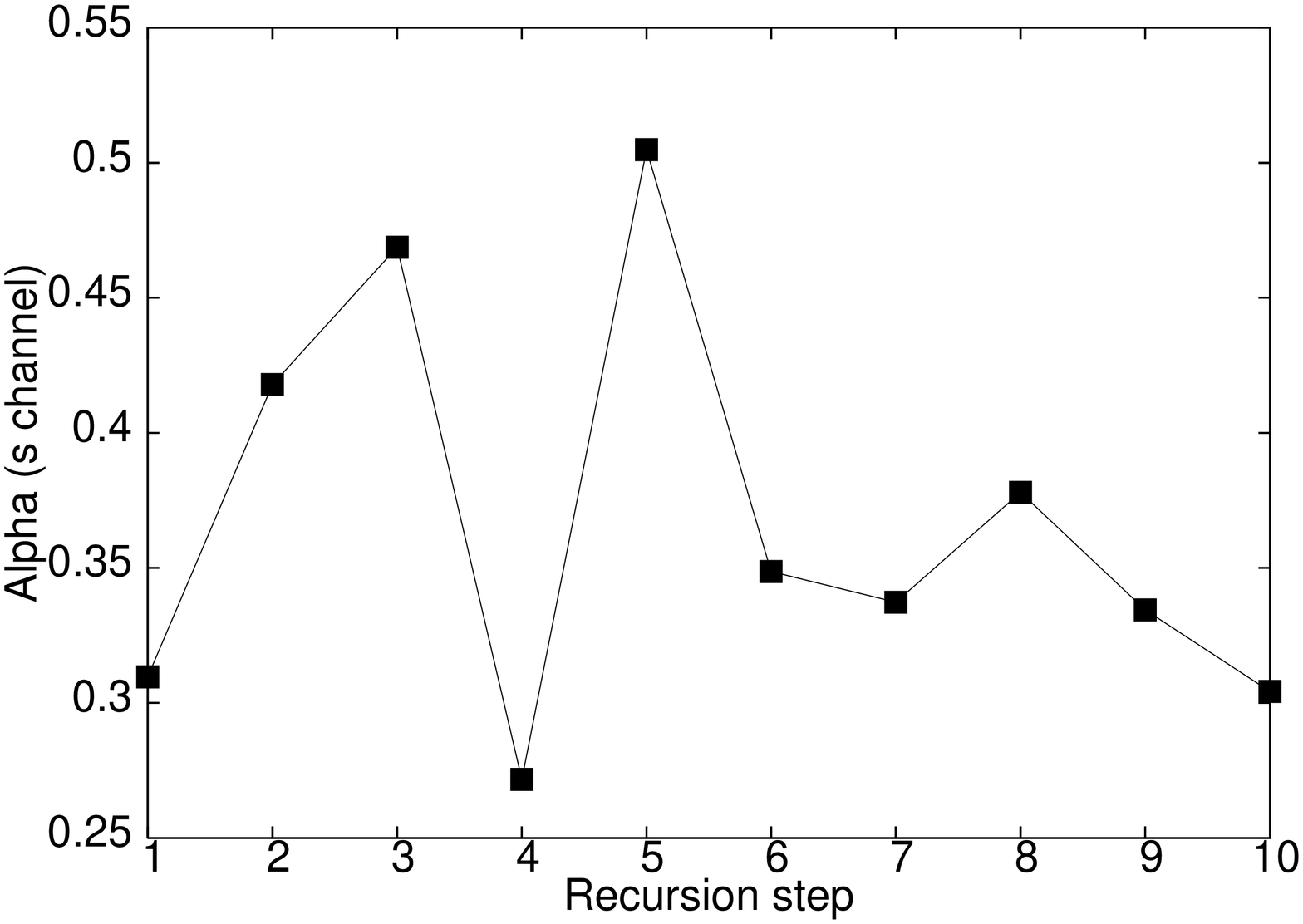}}
\epsfxsize=2in\epsfysize=2.5in\rotatebox{270}{\epsfbox{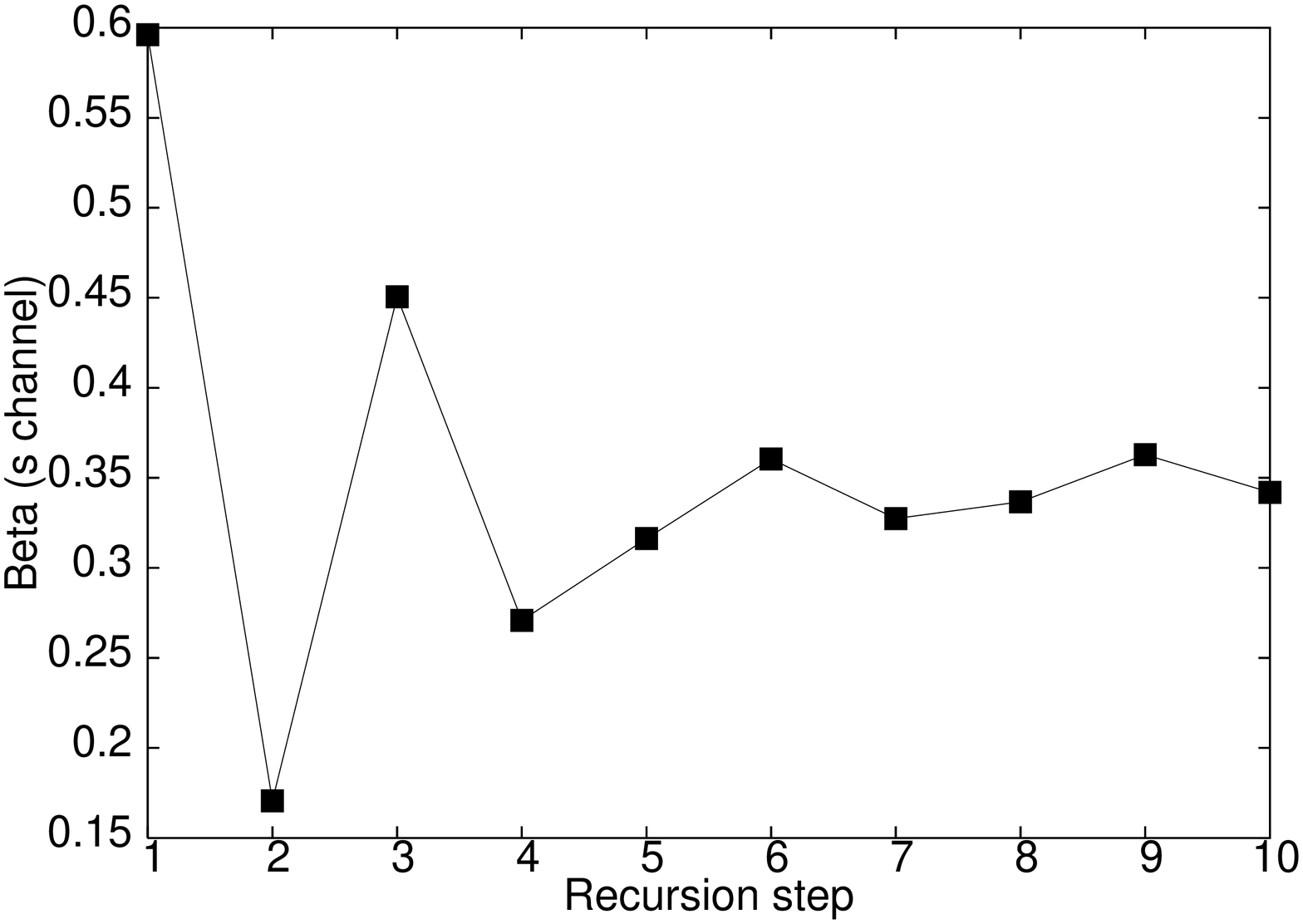}}\\
\epsfxsize=2in\epsfysize=2.5in\rotatebox{270}{\epsfbox{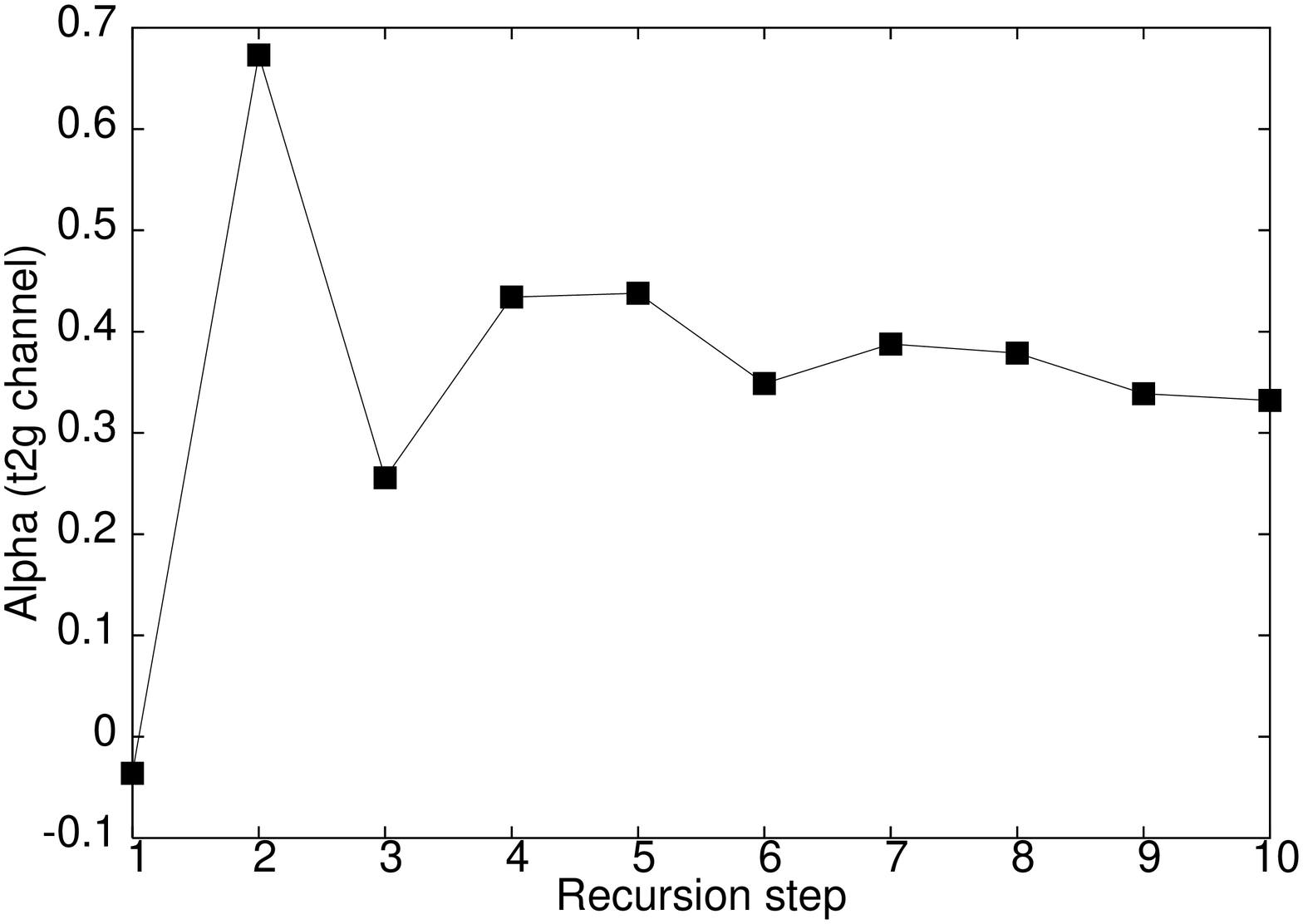}}
\epsfxsize=2in\epsfysize=2.5in\rotatebox{270}{\epsfbox{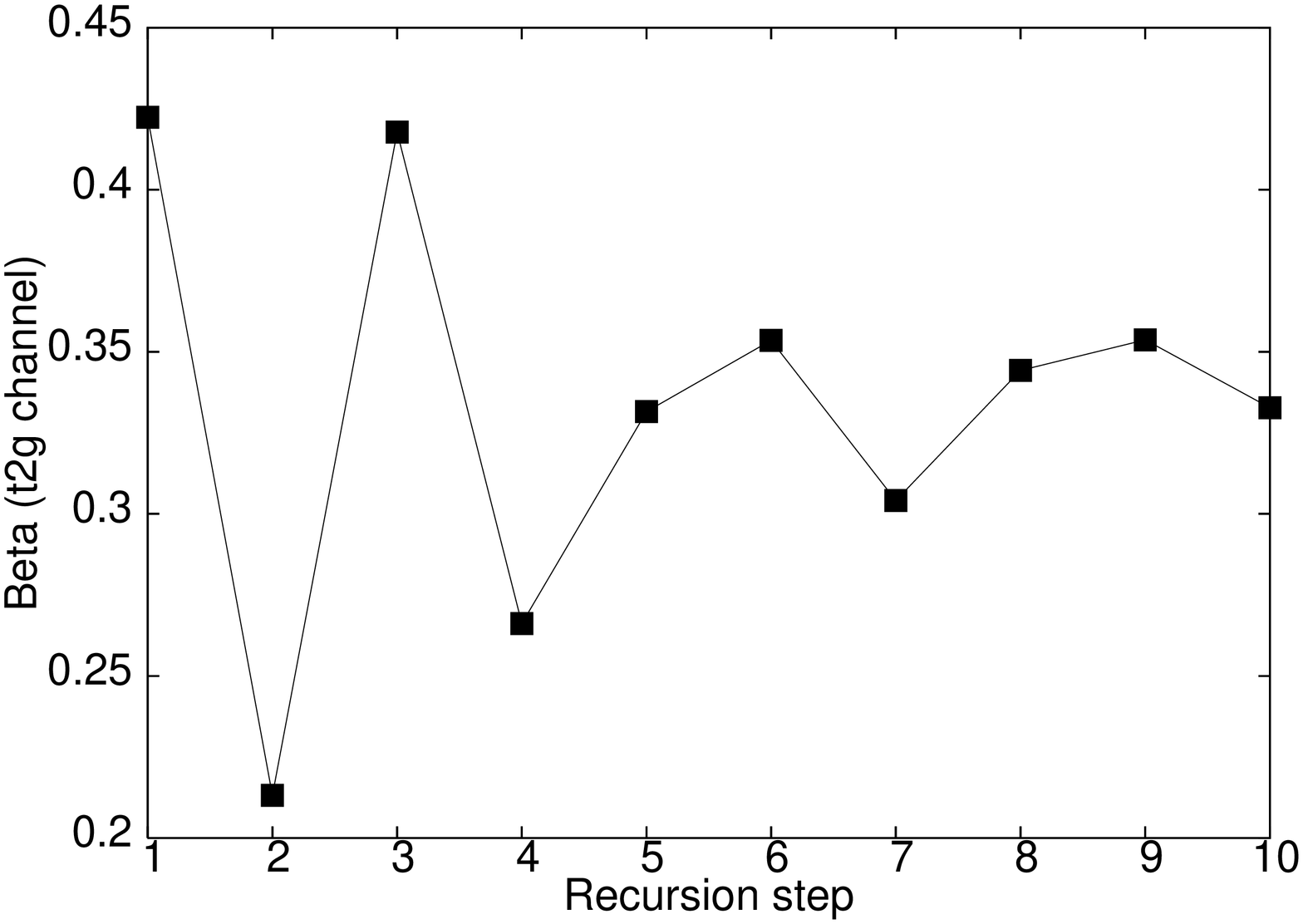}}\\
\caption{Continued fraction coefficients for the $s-$ and $t2g-$ channels for CuAu
alloy}
\label{fig2}
\end{figure}

In figure \ref{fig2} we show the continued fraction coefficients
for a few different angular momentum components. The coefficients  do show a tendency of showing oscillatory
 convergence,
but within the maximum ten steps that we were able to go, it was not possible to accurately locate
the converged values. We therefore chose not to use the square-root terminator, but opted for the
Beer-Pettifor terminator \cite{bp}.

\begin{figure}[h]
\centering
\vskip 1cm
\epsfxsize=3.5in \epsfysize=4in \rotatebox{270}{\epsfbox{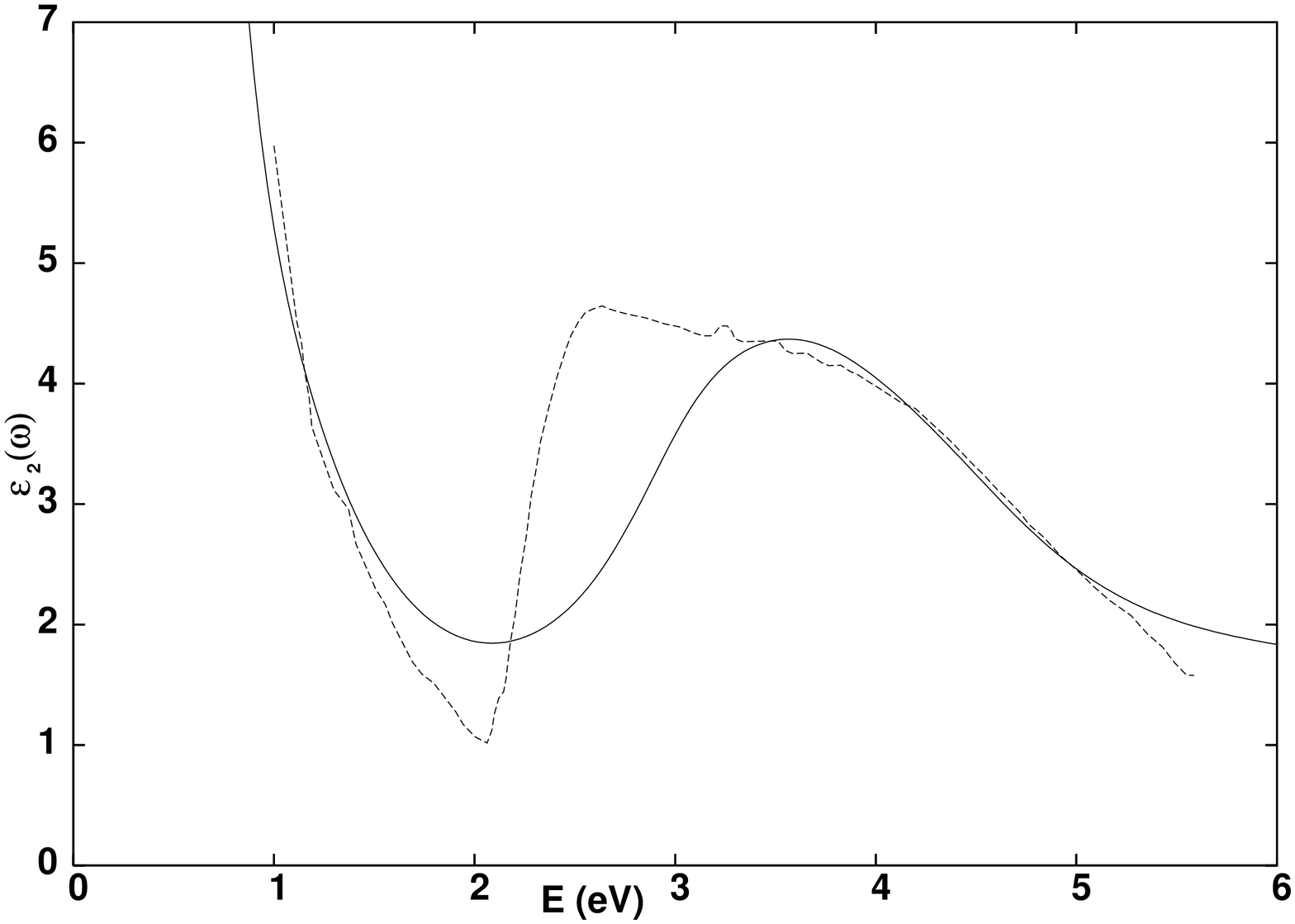}}
\caption{Averaged optical conductivity showing a Drude-like behaviour at low photon energies. Theoretical estimates are shown in full lines and 
experimental data for $\epsilon_2(\omega)$ in disordered CuAu taken from Scott and Muldawer \cite{sm} are shown in dotted lines}
\label{fig3}
\end{figure}

Experimental data on disordered CuAu (50-50) are available \cite{sm}. The authors
have reported high resolution optical data for both ordered and disordered CuAu for
photon energies between 1.2 eV and 6 eV.  In the
figure \ref{fig3},   we also show the experimental data for $\epsilon_2(\omega)$ in
dotted lines and compare it with the theoretical data in figure \ref{fig3}(bold lines). Drude
like behaviour is clearly seen till photon energies of $\sim$ 2.1 eV in theory
and around 2 eV in the experiment. Subsequently $\epsilon_2(\omega)$ rises again because of transitions from the $d$-states. Both the theory and experiment show a decrease around 6 eV. The experimental curve shows a sharper dip and the subsequent peak is at a lower energy. 

We have experimental data on AgAu (50-50) \cite{ban}, whose density of states closely resembles CuAu.
The inter-band contribution to the imaginary part of the dielectric function $\epsilon_2^\prime(\omega)$ may be
obtained from the optical conductivity data, by subtracting away the Drude contribution and dividing the result by $\omega$ : $\epsilon_2^\prime(\omega) = (\sigma(\omega)-\sigma^D(\omega))/\omega$ . Below the onset of the
inter-band transitions, this quantity vanishes. It shows a shoulder around 2.5 eV and then reaches a maximum at around 3.5 eV before decreasing and showing a feature around 5.5 eV.

\begin{figure}[h]
\centering
\vskip 1cm
\epsfxsize=3.5in \epsfysize=4in \rotatebox{270}{\epsfbox{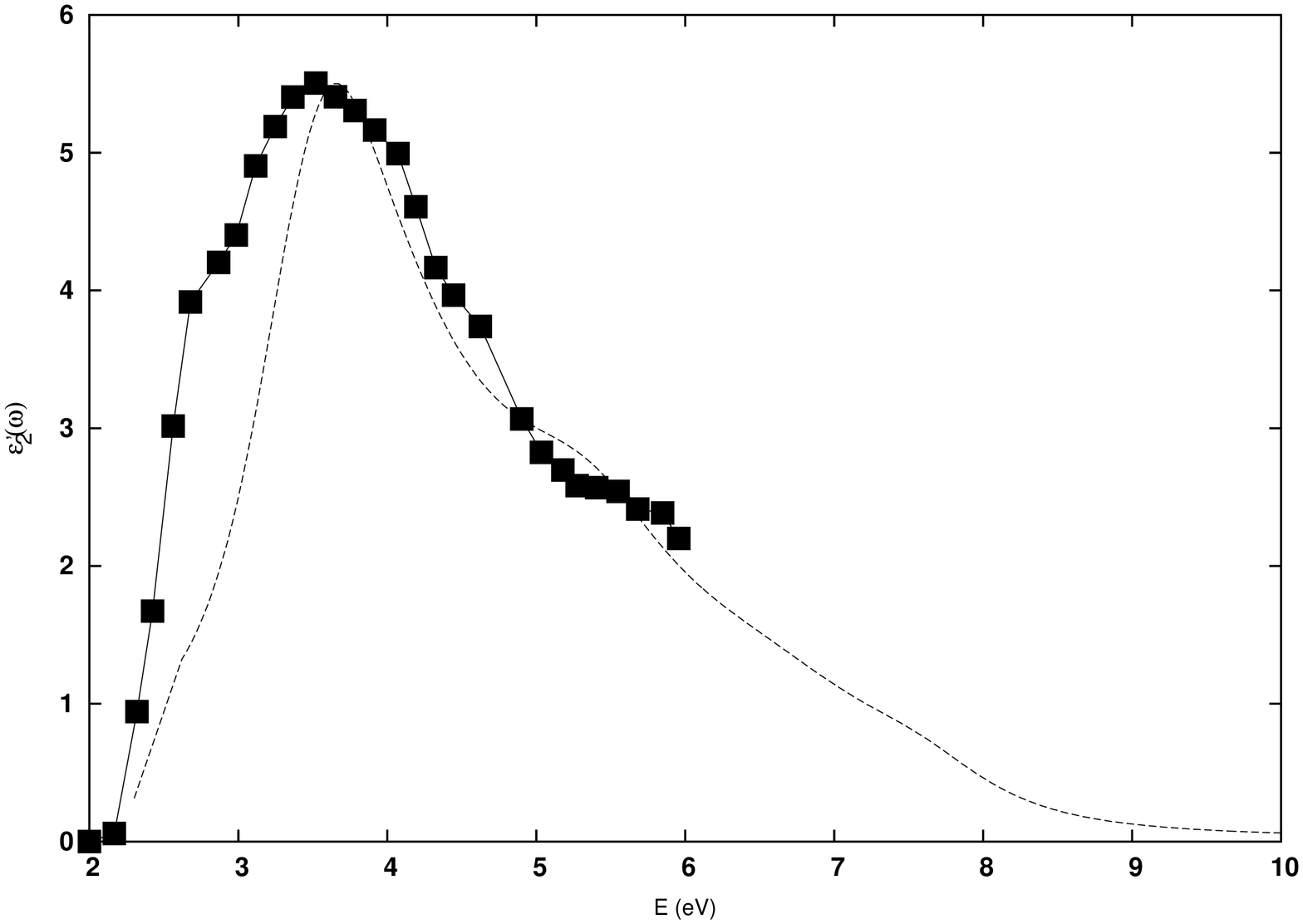}}
\caption{Intraband contribution to $\epsilon_2^\prime(\omega)$ for AgAu alloy  theoretical (full lines) and experimental (Nielsson \cite{niel}) (points)}
\label{fignew}
\end{figure}

As in the coherent potential approximation (CPA) estimates of Banhart \cite{ban} there is a $\simeq$ 0.5eV shift outwards in the theoretical estimate. In the above figure we have shifted our theoretical data by 0.5eV downwards for comparison.
Banhart attributes this shift to the foundations of the electronic theory used : local density approximation which cannot reproduce excited states properly. This had been discussed also by Fehrenbach and Bross \cite{fb}. To this we shall add that it is doubtful whether the energy linearized TB-LMTO can accurately reproduce the large energy window of our calculation. It may be useful in future to base our
calculations on the newly developed NMTO formalism \cite{sa}. In addition, the termination procedure of
the recursion requires further refinement, as the convergence of the continued fraction coefficients is
not as rapid as in density of states calculations. Banhart notes that the CPA neglects all short-ranged order. However, since our calculations in the augmented space recursion does include correlated scattering from different sites, we cannot attribute the discrepancy to the CPA.

Apart from the energy shift, the experimental data are in reasonable agreement with  our theory. 
The general shape with  a maximum is clearly reproduced. The feature shown in the
experimental data around 5.5 eV is also seen in our theoretical estimate, slightly shifted in energy.
The shoulder in the experimental data just above 2 eV is not properly reproduced in the theoretical
curve. The CPA estimate of Banhart (\cite{ban}) also does not seem to show such a prominent shoulder.
Is it possible that the experimental alloy had short ranged order ? It is known that short-ranged
order can have influence on optical properties (\cite{aah}). The augmented state formalism can handle
short-ranged order (\cite{sro}) and we shall study this effect in a future communication.

\ack One of us (KT) would like to acknowledge financial support from the CSIR, India.

\end{document}